\documentclass[preprint,showkeys,letterpaper]{revtex4}

\usepackage{amsfonts,amsmath,amssymb}
\usepackage{bm}
\usepackage[dvips]{graphicx}
\usepackage{hyperref}

\begin{document}

\title{The Extended Cartan Homotopy Formula and a\\Subspace Separation Method for Chern--Simons Theory}

\author{Fernando Izaurieta}
\email{fizaurie@gmail.com}

\author{Eduardo Rodr\'{\i}guez}
\email{edurodriguez@udec.cl}

\affiliation{Arnold Sommerfeld Center for Theoretical Physics,
Ludwig-Maximilians-Universit\"{a}t M\"{u}nchen, Theresienstra\ss e 37, D-80333 Munich, Germany}

\affiliation{Departament de F\'{\i}sica Te\`{o}rica, Universitat de Val\`{e}ncia,
46100 Burjassot, Val\`{e}ncia, Spain}

\affiliation{Departamento de F\'{\i}sica,
Universidad de Concepci\'{o}n, Casilla 160-C, Concepci\'{o}n, Chile}

\author{Patricio Salgado}
\email{pasalgad@udec.cl}

\affiliation{Departament de F\'{\i}sica Te\`{o}rica, Universitat de Val\`{e}ncia,
46100 Burjassot, Val\`{e}ncia, Spain}

\affiliation{Departamento de F\'{\i}sica,
Universidad de Concepci\'{o}n, Casilla 160-C, Concepci\'{o}n, Chile}

\date{March 8, 2006}

\preprint{LMU-ASC 14/06}

\begin{abstract}
In the context of Chern--Simons (CS) Theory, a \emph{subspace separation method} for the Lagrangian is proposed. The method is based on the iterative use of the Extended Cartan Homotopy Formula, and allows one to (1) separate the action in bulk and boundary contributions, and (2) systematically split the Lagrangian in appropriate reflection of the the subspace structure of the gauge algebra. In order to apply the method, one must regard CS forms as a particular case of more general objects known as \emph{transgression forms}. Five-dimensional CS Supergravity is used as an example to illustrate the method.
\end{abstract}


\maketitle

\section{\label{intro}Introduction}

Chern--Simons theory has been the subject of much interest in past decades. Its relevance for the formulation of higher-dimensional Supergravity models has been highlighted in, e.g.,~\cite{Cha90,Banh96,Tro97,Hor97,Tro98,Nas03,Iza06c}. It has even been suggested that they could provide with an action for M~Theory in eleven dimensions~\cite{Tro97,Hor97,Nas03}.

Since superalgebras come naturally split into distinct subspaces, it becomes essential, in order to get a taste of the theory's physical content, to be able to let this structure be felt on the Lagrangian itself. In other words, one must have a way to separate the Lagrangian into pieces that reflect the inner subspace structure of the gauge superalgebra.

In dimensions higher than three, the CS form becomes a nonlinear function of the gauge potential $\bm{A}$ and its associated field strength $\bm{F}$. This nonlinearity and the complexities of the superalgebras themselves make for a huge difficulty when it comes to performing the above-mentioned subspace separation.

The main aim of this paper is to present a method, based on the iterative use of the Extended Cartan Homotopy Formula (ECHF)~\cite{Man85}, which greatly eases the accomplishment of two important tasks. First, it allows the separation of the CS action into bulk and boundary contributions. Second, and perhaps most importantly, it permits the splitting of the bulk Lagrangian into pieces that reflect the particular subspace structure of the gauge algebra.

The formulation of the method requires regarding CS forms as particular cases of more general objects known as \emph{transgression forms}. The Transgression Form is the matrix where CS forms stem from~\cite{deAz95,Nak03}. It depends on \emph{two} one-form gauge connections $\bm{A}$ and $\bar{\bm{A}}$ and features prominently on the Chern--Weil Theorem~\cite{deAz95}, which expresses its exterior derivative as the difference of the Chern characters corresponding to both connections. A crucial feature of the transgression form is its full invariance under gauge transformations. The CS form corresponds to the case $\bar{\bm{A}}=0$; its well-known pseudo-gauge invariance can be traced back to the fact that this fixing is not a gauge-invariant one~\footnote{By the pseudo-gauge invariance of the CS form we mean the fact that it changes by a closed form under gauge transformations.}.

The paper is organized as follows. In Section~\ref{tgft} some general properties of transgression forms are reviewed. Section~\ref{s:echf} presents the ECHF and shows how a subspace separation method that allows for a deeper understanding of the CS Lagrangian can be built upon it. Five-dimen\-sional CS supergravity (SUGRA) is recalled in Section~\ref{su5} as an example of the use of the Method within the Transgression/CS framework. We close with conclusions and final remarks in Section~\ref{last}.

\section{\label{tgft}Chern--Simons Theory and Transgression Forms}

Let $\mathfrak{g}$ be a Lie Algebra~\footnote{Superalgebras can be considered as well, and we do so in Section~\ref{su5}, but we shall restrict ourselves to bosonic Lie Algebras for the moment in order to focus attention on the essential features of the theory.} and let us consider the theory defined on an orientable, $\left( 2n+1 \right)$-dimensional space-time manifold $M$ by the action~\cite{Bor03,Mor04a,Fat04,Fat05,Bor05,Iza05,Mor05,Mor06a,Mor06b}
\begin{equation}
S_{\mathrm{T}}^{\left( 2n+1 \right)} \left[ \bm{A}, \bar{\bm{A}} \right] =
k \int_{M} Q_{\bm{A} \leftarrow \bar{\bm{A}}}^{\left( 2n+1 \right)},
\label{st}
\end{equation}
where $k$ is a constant and the $\left( 2n+1 \right)$-form $Q_{\bm{A} \leftarrow \bar{\bm{A}}}^{\left( 2n+1 \right)}$ is the so-called \emph{transgression form}, defined by
\begin{equation}
Q_{\bm{A} \leftarrow \bar{\bm{A}}}^{\left( 2n+1 \right)} =
\left( n+1 \right) \int_{0}^{1} dt \left\langle \bm{\theta F}_{t}^{n} \right\rangle .
\end{equation}
Here $\bm{A}$ and $\bar{\bm{A}}$ are two $\mathfrak{g}$-valued, one-form gauge connections, and
\begin{eqnarray}
\bm{\theta} & = & \bm{A} - \bar{\bm{A}}, \label{theta} \\
\bm{A}_{t} & = & \bar{\bm{A}} + t \bm{\theta}, \\
\bm{F}_{t} & = & \mathrm{d}\bm{A}_{t} + \bm{A}_{t}^{2}.
\end{eqnarray}
The brackets $\left\langle \cdots \right\rangle$ stand for a rank $n+1$, symmetric $\mathfrak{g}$-invariant polynomial,
\begin{equation}
\left\langle \cdots \right\rangle : \; \underset{n+1}{\underbrace {\mathfrak{g} \times \cdots \times \mathfrak{g}}} \; \rightarrow \mathbb{R}.
\end{equation}
The choice, and the very existence of this bracket are all too important features that shape the theory to a great extent. Here it will suffice to remark that, given an explicit matrix representation for $\mathfrak{g}$, it is always possible to find such a polynomial~\footnote{For certain classes of algebras, including the M~Algebra, one can go beyond the supertrace and consider more general alternatives for this polynomial~\cite{Iza06c,Iza06b}.} (see Ref.~\cite{deAz97}). Clearly, the CS action can be regarded as the particular case $\bar{\bm{A}}=0$.

The action~(\ref{st}) is invariant under two large sets of independent symmetries. First, it is trivially invariant under diffeomorphisms, since it is constructed out of differential forms. Second, it is invariant under the (in general) non-abelian, local gauge transformations
\begin{eqnarray}
\bm{A} \rightarrow \bm{A}^{\prime}
& = &
g \left( \bm{A} - g^{-1} \mathrm{d} g \right) g^{-1},
\label{da1} \\
\bar{\bm{A}} \rightarrow \bar{\bm{A}}^{\prime}
& = &
g \left( \bar{\bm{A}} - g^{-1} \mathrm{d} g \right) g^{-1},
\label{da2}
\end{eqnarray}
with $g \left( x \right) = \exp \left( \lambda^{A} \left( x \right) \bm{G}_{A} \right)$, where $\left\{ \bm{G}_{A}, A = 1, \ldots, \dim \left( \mathfrak{g} \right) \right\}$ is a basis for $\mathfrak{g}$. The connections $\bm{A}$ and $\bar{\bm{A}}$ are regarded here (and throughout this work) as local one-form on spacetime rather than as global objects on the corresponding fiber-bundle, thus accounting for the transformation laws~(\ref{da1})--(\ref{da2}). From these transformation laws one can readily check that $\bm{\theta}$, defined in Equation~(\ref{theta}) as the difference between both connections, transforms as a tensor, i.e.
\begin{equation}
\bm{\theta} \rightarrow \bm{\theta}^{\prime} = g \bm{\theta} g^{-1}.
\end{equation}
In terms of a fiber-bundle description, these different behaviors correspond to the fact that the global fiber-bundle connections are not projectable to the base manifold whereas their difference is (see, e.g., Refs.~\cite{deAz95,Bor05}).

The full invariance of the action~(\ref{st}) under~(\ref{da1})--(\ref{da2}) rests on the invariance property of the symmetric polynomial $\left\langle \cdots \right\rangle$ and on the tensor transformation laws for both $\bm{\theta}$ and $\bm{F}_{t}$. It is also deeply related to the Chern--Weil Theorem, see \cite{deAz95,Nak03,Iza05}.

In the CS case, where $\bar{\bm{A}}=0$ from the outset, this full invariance is reduced to pseudo-invariance. This is a consequence of the fact that the fixing $\bar{\bm{A}}=0$ is not a gauge-invariant one (see, e.g., Refs.~\cite{deAz95,Iza05,Mor05}).

However straightforward to establish, both the diffeomorphism and the gauge symmetries are far-reaching, as is proved by the fact that they lead to nontrivial conserved charges~\cite{Iza05,Mor05}.

The field equations for the transgression action~(\ref{st}) read
\begin{eqnarray}
\left\langle \bm{F}^{n} \bm{G}_{A} \right\rangle & = & 0, \label{Fn=0} \\
\left\langle \bar{\bm{F}}^{n} \bm{G}_{A} \right\rangle & = & 0,
\end{eqnarray}
and the boundary conditions are
\begin{equation}
\left. \int_{0}^{1} dt \left\langle \delta \bm{A}_{t} \bm{\theta F}_{t}^{n-1} \right\rangle \right\vert_{\partial M} = 0. \label{bc}
\end{equation}
A deeper analysis of the physics produced by the transgression Lagrangian can be found in~\cite{Iza05,Mor05}. For the CS case, only Equation~(\ref{Fn=0}) is present.

\section{\label{s:echf}The ECHF and the CS Subspace Separation Method}

As made apparent in Section~\ref{tgft}, the CS and transgression Lagrangians allow by themselves for a general understanding of the formal aspects of the theory. The detailed physical behavior, however, is highly dependent on the choice of gauge group and invariant polynomial. Since gauge groups used in practice often have distinct physically meaningful subspaces, it is useful to write the Lagrangian in such a way as to reflect this structure.

The following sections outline a Subspace Separation Method for CS and Transgression Lagrangians. The method is based on a particular case of the Extended Cartan Homotopy Formula, which is reviewed for completeness in Section~\ref{sub:echf}.

\subsection{\label{sub:echf}The Extended Cartan Homotopy Formula}

\begin{figure}
\centerline{\includegraphics[width=.8\columnwidth]{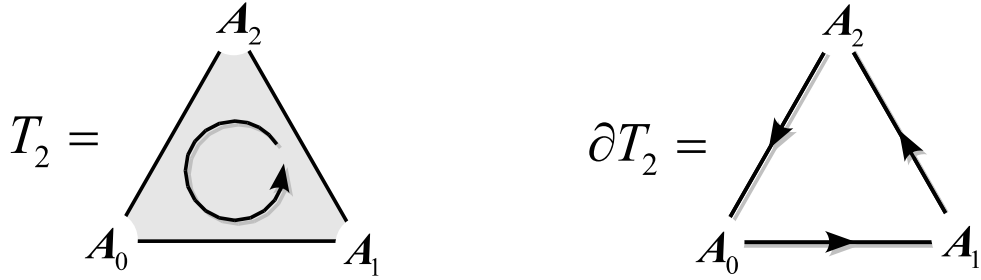}}
\caption{A two-dimensional simplex $T_{2} = \left( \bm{A}_{0} \bm{A}_{1} \bm{A}_{2} \right)$ and its boundary, $\partial T_{2} = \left( \bm{A}_{1} \bm{A}_{2} \right) - \left( \bm{A}_{0} \bm{A}_{2} \right) + \left( \bm{A}_{0} \bm{A}_{1} \right)$. The one-form gauge connections $\bm{A}_{0}$, $\bm{A}_{1}$ and $\bm{A}_{2}$ are pictured as associated to the simplex vertices.}
\label{mono}
\end{figure}

Let us consider a set $\left\{ \bm{A}_{i}, i = 0, \ldots, r+1 \right\} $ of one-form gauge connections on a fiber-bundle over a $d$-dimensional manifold $M$ and a $\left( r+1 \right)$-dimensional oriented simplex $T_{r+1}$ parametrized by the set $\left\{ t^{i}, i = 0, \ldots, r+1 \right\}$. These parameters must satisfy the constraints
\begin{eqnarray}
t^{i} & \geq & 0, \qquad i = 0, \ldots, r+1, \\
\sum_{i=0}^{r+1} t^{i} & = & 1. \label{tt1}
\end{eqnarray}
Equation~(\ref{tt1}) in particular implies that the linear combination
\begin{equation}
\bm{A}_{t} = \sum_{i=0}^{r+1} t^{i} \bm{A}_{i}
\end{equation}
transforms as a gauge connection in the same way as every individual $\bm{A}_{i}$ does. We can picture each $\bm{A}_{i}$ as associated to the $i$-th vertex of $T_{r+1}$ (see Figure~\ref{mono}), which we accordingly denote as
\begin{equation}
T_{r+1} = \left( \bm{A}_{0} \bm{A}_{1} \cdots \bm{A}_{r+1} \right).
\end{equation}

With the preceding notation, the ECHF reads~\cite{Man85}
\begin{equation}
\int_{\partial T_{r+1}} \frac{l_{t}^{p}}{p!} \pi = \int_{T_{r+1}} \frac{l_{t}^{p+1}} {\left( p+1 \right)!} \mathrm{d} \pi + \left( -1 \right)^{p+q} \mathrm{d} \int_{T_{r+1}} \frac{l_{t}^{p+1}}{\left( p+1 \right)!} \pi. \label{cehf}
\end{equation}
Here $\pi$ represents a polynomial in the forms $\left\{ \bm{A}_{t}, \bm{F}_{t}, \mathrm{d}_{t} \bm{A}_{t}, \mathrm{d}_{t} \bm{F}_{t} \right\}$ which is also an $m$-form on $M$ and a $q$-form on $T_{r+1}$, with $m \geq p$ and $p+q=r$. The exterior derivatives on $M$ and $T_{r+1}$ are denoted respectively by $\mathrm{d}$ and $\mathrm{d}_{t}$. The operator $l_{t}$, called \textit{homotopy derivation}, maps differential forms on $M$ and $T_{r+1}$ according to
\begin{equation}
l_{t} : \Omega^{a} \left( M \right) \times \Omega^{b} \left( T_{r+1} \right) \rightarrow \Omega^{a-1} \left( M \right) \times \Omega^{b+1} \left( T_{r+1} \right),
\end{equation}
and it satisfies Leibniz's rule together with $\mathrm{d}$ and $\mathrm{d}_{t}$. Its action on $\bm{A}_{t}$ and $\bm{F}_{t}$ reads~\cite{Man85}
\begin{eqnarray}
l_{t} \bm{F}_{t} & = & \mathrm{d}_{t} \bm{A}_{t}, \label{lfda} \\
l_{t} \bm{A}_{t} & = & 0.
\end{eqnarray}

The three operators $\mathrm{d}$, $\mathrm{d}_{t}$ and $l_{t}$ define a graded algebra given by
\begin{eqnarray}
\mathrm{d}^{2} & = & 0, \label{ga1} \\
\mathrm{d}_{t}^{2} & = & 0, \\
\left[ l_{t}, \mathrm{d} \right] & = & \mathrm{d}_{t}, \\
\left[ l_{t}, \mathrm{d}_{t} \right] & = & 0, \\
\left\{ \mathrm{d}, \mathrm{d}_{t} \right\} & = & 0. \label{ga5}
\end{eqnarray}

Let us now pick the following polynomial:
\begin{equation}
\pi = \left\langle \bm{F}_{t}^{n+1} \right\rangle.
\end{equation}
This choice has the three following properties: (1) $\pi$ is $M$-closed~\footnote{This is easily deduced from the invariant property of $\pi$ and Bianchi's identity D$_{t}\bm{F}_{t}=0$.}, i.e., $\mathrm{d} \pi = 0$, (2) $\pi$ is a 0-form on $T_{r+1}$, i.e., $q = 0$ and (3) $\pi$ is a $\left( 2n+2 \right)$-form on $M$, i.e., $m = 2n+2$. The allowed values for $p$ are $p = 0, \ldots, 2n+2$. The ECHF reduces in this case to
\begin{equation}
\int_{\partial T_{p+1}} \frac{l_{t}^{p}}{p!} \left\langle \bm{F}_{t}^{n+1} \right\rangle = \left( -1 \right)^{p} \mathrm{d} \int_{T_{p+1}} \frac{l_{t}^{p+1}}{\left( p+1 \right)!} \left\langle \bm{F}_{t}^{n+1} \right\rangle. \label{cehfc}
\end{equation}
We call Equation~(\ref{cehfc}) the `restricted' (or `closed') version of the ECHF.

A first well-known particular case of the ECHF is the Chern--Weil theorem. Setting $p=0$ in Equation~(\ref{cehfc}) one readily finds
\begin{equation}
\left\langle \bm{F}_{1}^{n+1} \right\rangle - \left\langle \bm{F}_{0}^{n+1} \right\rangle = \mathrm{d} Q_{\bm{A}_{1} \leftarrow \bm{A}_{0}}^{\left( 2n+1 \right)},
\end{equation}
where the \emph{transgression form} $Q_{\bm{A}_{1} \leftarrow \bm{A}_{0}}^{\left( 2n+1 \right)}$ turns out to be defined by
\begin{eqnarray}
Q_{\bm{A}_{1} \leftarrow \bm{A}_{0}}^{\left( 2n+1 \right)} & \equiv & \int_{T_{1}} l_{t} \left\langle \bm{F}_{t}^{n+1} \right\rangle \nonumber \\
& = & \left( n+1 \right) \int_{0}^{1} dt \left\langle \left( \bm{A}_{1} - \bm{A}_{0} \right) \bm{F}_{t}^{n} \right\rangle.
\end{eqnarray}

A second particular case which is directly relevant in the present context corresponds to setting $p = 1$ in Equation~(\ref{cehfc}),
\begin{equation}
\int_{\partial T_{2}} l_{t} \left\langle \bm{F}_{t}^{n+1} \right\rangle = - \mathrm{d} \int_{T_{2}} \frac{l_{t}^{2}}{2} \left\langle \bm{F}_{t}^{n+1} \right\rangle, \label{p1}
\end{equation}
where $\bm{F}_{t}$ is the curvature corresponding to the connection $\bm{A}_{t} = t^{0} \bm{A}_{0} +t^{1} \bm{A}_{1} + t^{2} \bm{A}_{2}$. The boundary of the simplex $T_{2} = \left( \bm{A}_{0} \bm{A}_{1} \bm{A}_{2} \right)$ may be written as the sum (see Figure~\ref{mono})
\begin{equation}
\partial\left( \bm{A}_{0} \bm{A}_{1} \bm{A}_{2} \right) = \left( \bm{A}_{1} \bm{A}_{2} \right) - \left( \bm{A}_{0} \bm{A}_{2} \right) + \left( \bm{A}_{0} \bm{A}_{1} \right),
\end{equation}
so that the integral in the left-hand side of~(\ref{p1}) is decomposed as
\begin{equation}
\int_{\partial T_{2}} l_{t} \left\langle \bm{F}_{t}^{n+1} \right\rangle = \int_{\left( \bm{A}_{1} \bm{A}_{2} \right)} l_{t} \left\langle \bm{F}_{t}^{n+1} \right\rangle - \int_{\left( \bm{A}_{0} \bm{A}_{2} \right)} l_{t} \left\langle \bm{F}_{t}^{n+1} \right\rangle + \int_{\left( \bm{A}_{0} \bm{A}_{1} \right)} l_{t} \left\langle \bm{F}_{t}^{n+1} \right\rangle.
\end{equation}
Each of the terms in this equation corresponds to what was called before a transgression form:
\begin{equation}
\int_{\partial T_{2}} l_{t} \left\langle \bm{F}_{t}^{n+1} \right\rangle = Q_{\bm{A}_{2} \leftarrow \bm{A}_{1}}^{\left( 2n+1 \right)} - Q_{\bm{A}_{2} \leftarrow \bm{A}_{0}}^{\left( 2n+1 \right)} + Q_{\bm{A}_{1} \leftarrow \bm{A}_{0}}^{\left( 2n+1 \right)}.
\end{equation}

On the other hand, Leibniz's rule for $l_{t}$ and Equation~(\ref{lfda}) together imply that
\begin{equation}
\int_{T_{2}} \frac{l_{t}^{2}}{2} \left\langle \bm{F}_{t}^{n+1} \right\rangle = \frac{1}{2} n \left( n+1 \right) \int_{T_{2}} \left\langle \left( \mathrm{d}_{t} \bm{A}_{t} \right)^{2} \bm{F}_{t}^{n-1} \right\rangle.
\end{equation}
Integrating over the simplex one gets
\begin{equation}
\int_{T_{2}} \frac{l_{t}^{2}}{2} \left\langle \bm{F}_{t}^{n+1} \right\rangle = Q_{\bm{A}_{2} \leftarrow \bm{A}_{1} \leftarrow \bm{A}_{0}}^{\left( 2n \right)},
\end{equation}
where $Q_{\bm{A}_{2} \leftarrow \bm{A}_{1} \leftarrow \bm{A}_{0}}^{\left( 2n \right)}$ is given by
\begin{equation}
Q_{\bm{A}_{2} \leftarrow \bm{A}_{1} \leftarrow \bm{A}_{0}}^{\left( 2n \right)} \equiv n \left( n+1 \right) \int_{0}^{1} dt \int_{0}^{t} ds \left\langle \left( \bm{A}_{2} - \bm{A}_{1} \right) \left( \bm{A}_{1} - \bm{A}_{0} \right) \bm{F}_{t}^{n-1} \right\rangle. \label{q300}
\end{equation}
In~(\ref{q300}) we have introduced dummy parameters $t=1-t^{0}$ and $s=t^{2}$,
in terms of which $\bm{A}_{t}$ reads
\begin{equation}
\bm{A}_{t} = \bm{A}_{0} + s \left( \bm{A}_{2} - \bm{A}_{1} \right) + t \left( \bm{A}_{1} - \bm{A}_{0} \right).
\end{equation}

Putting everything together, we find the Triangle Equation
\begin{equation}
Q_{\bm{A}_{2} \leftarrow \bm{A}_{1}}^{\left( 2n+1 \right)} - Q_{\bm{A}_{2} \leftarrow \bm{A}_{0}}^{\left( 2n+1 \right)} + Q_{\bm{A}_{1} \leftarrow \bm{A}_{0}}^{\left( 2n+1 \right)} = - \mathrm{d} Q_{\bm{A}_{2} \leftarrow \bm{A}_{1} \leftarrow \bm{A}_{0}}^{\left( 2n \right)},
\end{equation}
or alternatively
\begin{equation}
Q_{\bm{A}_{2} \leftarrow \bm{A}_{0}}^{\left( 2n+1 \right)} = Q_{\bm{A}_{2} \leftarrow \bm{A}_{1}}^{\left( 2n+1 \right)} + Q_{\bm{A}_{1} \leftarrow \bm{A}_{0}}^{\left( 2n+1 \right)} + \mathrm{d} Q_{\bm{A}_{2} \leftarrow \bm{A}_{1} \leftarrow \bm{A}_{0}}^{\left( 2n \right)}. \label{treq0}
\end{equation}
We would like to stress here that use of the ECHF has allowed us to pinpoint the exact form of the boundary contribution $Q_{\bm{A}_{2} \leftarrow \bm{A}_{1} \leftarrow \bm{A}_{0}}^{\left( 2n \right)}$, Equation~(\ref{q300}).

\subsection{The Subspace Separation Method}

In Section~\ref{sub:echf} the ECHF was reviewed in detail and two particular cases were examined. In this section, a subspace separation method for CS Lagrangians is built upon the second of them, namely, the triangle equation:
\begin{equation}
Q_{\bm{A}_{2} \leftarrow \bm{A}_{0}}^{\left( 2n+1 \right)} = Q_{\bm{A}_{2} \leftarrow \bm{A}_{1}}^{\left( 2n+1 \right)} + Q_{\bm{A}_{1} \leftarrow \bm{A}_{0}}^{\left( 2n+1 \right)} + \mathrm{d} Q_{\bm{A}_{2}\leftarrow
\bm{A}_{1}\leftarrow\bm{A}_{0}}^{\left(  2n\right)  }. \label{treq}
\end{equation}
The triangle equation~(\ref{treq}) splits a transgression form $Q_{\bm{A}_{2} \leftarrow \bm{A}_{0}}^{\left( 2n+1 \right)}$ into the sum of two transgression forms depending on an `intermediate' connection $\bm{A}_{1}$ plus an exact form. The detailed form of $Q_{\bm{A}_{2} \leftarrow \bm{A}_{1} \leftarrow \bm{A}_{0}}^{\left( 2n \right)}$ reads~[cf.~Equation~(\ref{q300})]
\begin{equation}
Q_{\bm{A}_{2} \leftarrow \bm{A}_{1} \leftarrow \bm{A}_{0}}^{\left( 2n \right)} \equiv n \left( n+1 \right) \int_{0}^{1} dt \int_{0}^{t} ds \left\langle \left( \bm{A}_{2} - \bm{A}_{1} \right) \left( \bm{A}_{1} - \bm{A}_{0} \right) \bm{F}_{st}^{n-1} \right\rangle, \label{q3}
\end{equation}
where $\bm{F}_{st}$ is the curvature corresponding to the `interpolating'
connection
\begin{equation}
\bm{A}_{st} = \bm{A}_{0} + s \left( \bm{A}_{2} - \bm{A}_{1} \right) + t \left( \bm{A}_{1} - \bm{A}_{0} \right). \label{ast}
\end{equation}

It is worth stressing that, while each term in the right-hand side of~(\ref{treq}) depends on the intermediate connection $\bm{A}_{1}$, they do so in such a way that their sum depends solely on $\bm{A}_{2}$ and $\bm{A}_{0}$, matching what is found in the left-hand side.

The subspace separation method is based on the triangle equation~(\ref{treq}), and embodies the following steps:
\begin{enumerate}
\item Identify the relevant subspaces present in the gauge algebra, i.e., write $\mathfrak{g} = V_{0} \oplus \cdots \oplus V_{p}$.

\item Write the connections in terms of pieces valued on every subspace, i.e., $\bm{A} = \bm{a}_{0} + \cdots + \bm{a}_{p}$, $\bar{\bm{A}} = \bar{\bm{a}}_{0} + \cdots + \bar{\bm{a}}_{p}$.

\item Use Equation~(\ref{treq}) with
\begin{eqnarray}
\bm{A}_{0} & = & \bar{\bm{A}}, \\
\bm{A}_{1} & = & \bm{a}_{0} + \cdots + \bm{a}_{p-1},\\
\bm{A}_{2} & = & \bm{A}.
\end{eqnarray}

\item Repeat step 3 for the transgression $Q_{\bm{A}_{1} \leftarrow \bm{A}_{0}}$, etc.
\end{enumerate}

After performing these steps, one ends up with an equivalent expression for the transgression Lagrangian which has been separated in two different ways. First, the Lagrangian is split into bulk and boundary contributions. This is due to the fact that each use of Equation~(\ref{treq}) brings in a new boundary term. Second, each term in the bulk Lagrangian refers to a different subspace of the gauge algebra. This comes about because the difference $\bm{A}_{2} - \bm{A}_{1}$ is valued only on one particular subspace.

This method will be demonstrated with an example in Section~\ref{su5}.

\section{\label{su5}A Separation Example: Five-Dimensional CS SUGRA}

In order to highlight the way in which the subspace separation method sketched in Section~\ref{s:echf} is used in practice, in this section we apply it to a simple five-dimensional case.

The standard five-dimensional CS Supergravity~\cite{Cha90,Banh96,Tro98} uses the $\mathcal{N}$-extended AdS Superalgebra $\mathfrak{u} \left( 4|\mathcal{N} \right)$. This algebra is generated by $\bm{M}_{n}^{\phantom{n} m}$, $\bm{K}$, $\bm{P}_{a}$, $\bm{J}_{ab}$, $\bm{Q}_{i}^{\alpha}$, $\bar{\bm{Q}}_{\alpha}^{i}$, where the following physically meaningful subspaces are present:
\begin{enumerate}
\item a $\mathfrak{u} \left( \mathcal{N} \right)$ subalgebra, generated by $\bm{M}_{n}^{\phantom{n} m}$ and $\bm{K},$
\item an AdS subalgebra, generated by $\bm{P}_{a}$ and $\bm{J}_{ab}$,
\item and a fermionic subspace, generated by the Dirac spinors $\bm{Q}_{i}^{\alpha}$ and $\bar{\bm{Q}}_{\alpha}^{i}$.
\end{enumerate}

One may also go one step further and separate the $\mathfrak{u} \left( \mathcal{N} \right)$ subalgebra in $\mathfrak{su} \left( \mathcal{N} \right)$ plus an abelian part and the AdS subalgebra in the Lorentz Algebra plus AdS boosts.

An essential ingredient in this construction is the symmetric invariant polynomial $\left\langle \cdots \right\rangle$. In the present case this will be simply given by the supersymmetrized supertrace of the product of three supermatrices representing as many generators in $\mathfrak{u} \left( 4|\mathcal{N} \right)$. We use Dirac Matrices in $d = 5$ to represent the AdS generators, while for the rest we choose the adjoint representation~\cite{Iza05}.

The CS Lagrangian is given by
\begin{equation}
L_{\mathrm{sugra}}^{\left( 5 \right)} = k Q_{\bm{A}_{5} \leftarrow \bm{A}_{0}}^{\left( 5 \right)}, \label{lsu}
\end{equation}
where
\begin{eqnarray}
\bm{A}_{0} & = & 0, \\
\bm{A}_{5} & = & \bm{a} + \bm{b} + \bm{e} + \bm{\omega} + \bar{\bm{\psi}} - \bm{\psi},
\end{eqnarray}
with
\begin{eqnarray}
\bm{e} & = & e^{a} \bm{P}_{a}, \\
\bm{\omega} & = & \frac{1}{2} \omega^{ab} \bm{J}_{ab}, \\
\bm{a} & = & a_{\phantom{m} n}^{m} \bm{M}_{m}^{\phantom{m} n}, \\
\bm{b} & = & b \bm{K}, \\
\bar{\bm{\psi}} & = & \bar{\psi}_{\alpha}^{k} \bm{Q}_{k}^{\alpha}, \\
\bm{\psi} & = & \bar{\bm{Q}}_{\alpha}^{k} \psi_{k}^{\alpha}.
\end{eqnarray}

In order to apply the subspace separation method to the Lagrangian~(\ref{lsu}) we introduce the following set of intermediate connections:
\begin{eqnarray}
\bm{A}_{1} & = & \bm{\omega}, \\
\bm{A}_{2} & = & \bm{e} + \bm{\omega}, \\
\bm{A}_{3} & = & \bm{b} + \bm{e} + \bm{\omega}, \\
\bm{A}_{4} & = & \bm{a} + \bm{b} + \bm{e} + \bm{\omega}.
\end{eqnarray}
Repeated use of the triangle equation~(\ref{treq}) now allows us to split the Lagrangian~(\ref{lsu}) as
\begin{equation}
L_{\mathrm{sugra}}^{\left( 5 \right)} = L_{\bm{\psi}}^{\left( 5 \right)} + L_{\bm{a}}^{\left( 5 \right)} + L_{\bm{b}}^{\left( 5 \right)} + L_{\bm{e}}^{\left( 5 \right)} + \mathrm{d} B_{\mathrm{sugra}}^{\left( 4 \right)},
\end{equation}
where
\begin{eqnarray}
L_{\bm{\psi}}^{\left( 5 \right)} & = & k Q_{\bm{A}_{5} \leftarrow \bm{A}_{4}}^{\left( 5 \right)}, \label{L54} \\
L_{\bm{a}}^{\left( 5 \right)} & = & k Q_{\bm{A}_{4} \leftarrow \bm{A}_{3}}^{\left( 5 \right)}, \label{L43} \\
L_{\bm{b}}^{\left( 5 \right)} & = & k Q_{\bm{A}_{3} \leftarrow \bm{A}_{2}}^{\left( 5 \right)}, \label{L32} \\
L_{\bm{e}}^{\left( 5 \right)} & = & k Q_{\bm{A}_{2} \leftarrow \bm{A}_{1}}^{\left( 5 \right) }, \label{L21}
\end{eqnarray}
and
\begin{equation}
B_{\mathrm{sugra}}^{\left( 4 \right)} = k Q_{\bm{A}_{5} \leftarrow \bm{A}_{4} \leftarrow \bm{A}_{0}}^{\left( 4 \right)} + k Q_{\bm{A}_{4} \leftarrow \bm{A}_{3} \leftarrow \bm{A}_{0}}^{\left( 4 \right)} + k Q_{\bm{A}_{3} \leftarrow \bm{A}_{2} \leftarrow \bm{A}_{0}}^{\left( 4 \right)} + k Q_{\bm{A}_{2} \leftarrow \bm{A}_{1} \leftarrow \bm{A}_{0}}^{\left( 4 \right)}. \label{Bsugra}
\end{equation}
A few comments are in order. Ignoring for the moment the boundary contribution $B_{\mathrm{sugra}}^{\left( 4 \right)}$, we see that all dependence on the fermions has been packaged in $L_{\bm{\psi}}^{\left( 5 \right)}$, which we call `fermionic Lagrangian'. Similarly, $L_{\bm{a}}^{\left( 5 \right)}$ and $L_{\bm{b}}^{\left( 5 \right)}$ correspond to pieces that are highly dependent on $\bm{a}$ and $\bm{b}$ respectively, although some dependence on $\bm{a}$ and $\bm{b}$ is also found on $L_{\bm{\psi}}^{\left( 5 \right)}$. In turn, $L_{\bm{b}}^{\left( 5 \right)}$ carries no dependence on $\bm{a}$. The last piece, $L_{\bm{e}}^{\left( 5 \right)}$, corresponds to an Lanczos--Lovelock-type Lagrangian for gravity~\cite{Lov71}.

Explicit versions for every piece may be easily obtained by going back to the definition of a transgression form. This amounts to a huge difference with the case where there is no tool to perform the separation. Using only the CS form as a starting point, calculations can quickly become intractable for dimensions higher than three. As a matter of fact, a straightforward computation gives
\begin{eqnarray}
L_{\bm{\psi}}^{\left( 5 \right)} & = & \frac{3k}{2i} \left( \bar{\psi} \mathcal{R} \nabla \psi + \bar{\psi}^{n} \mathcal{F}_{\phantom{m} n}^{m} \nabla \psi_{m} - \nabla \bar{\psi} \mathcal{R} \psi - \nabla \bar{\psi}^{n} \mathcal{F}_{\phantom{m} n}^{m} \psi_{m} \right), \\
L_{\bm{a}}^{\left( 5 \right)} & = & \frac{3k}{\mathcal{N}} \left( \mathrm{d} b \right) \mathrm{Tr} \left( a \mathrm{d} a + \frac{2}{3} a^{3} \right) - i k \mathrm{Tr} \left[ a \left( \mathrm{d} a \right)^{2} + \frac{3}{2} a^{3} \mathrm{d} a + \frac{3}{5} a^{5} \right], \\
L_{\bm{b}}^{\left( 5 \right)} & = & k \left( \frac{1}{4^{2}} - \frac{1}{\mathcal{N}^{2}} \right) b \left( \mathrm{d} b \right)^{2} - \frac{3k}{4 \ell^{2}} b \left( T^{a} T_{a} - R_{ab} e^{a} e^{b} - \frac{\ell^{2}}{2} R^{ab} R_{ab} \right), \\
L_{\bm{e}}^{\left( 5 \right)} & = & \frac{3k}{8 \ell} \varepsilon_{abcde} \left( R^{ab} R^{cd} + \frac{2}{3} R^{ab} e^{c} e^{d} + \frac{1}{5} e^{a} e^{b} e^{c} e^{d} \right) e^{e},
\end{eqnarray}
where
\begin{eqnarray}
\mathcal{R} & = & i \left( \frac{1}{4} + \frac{1}{\mathcal{N}} \right) \left( \mathrm{d} b + \frac{i}{2 \ell} \bar{\psi} \psi \right) \openone + \frac{1}{2} \left( T^{a} - \frac{1}{4} \bar{\psi} \Gamma^{a} \psi \right) \Gamma_{a} + \nonumber \\
& & + \frac{1}{4} \left( R^{ab} + \frac{1}{\ell^{2}} e^{a} e^{b} + \frac{1}{4 \ell} \bar{\psi} \Gamma^{ab} \psi\right) \Gamma_{ab}, \\
\mathcal{F}_{\phantom{m} n}^{m} & = & f_{\phantom{m} n}^{m} - \frac{1}{2 \ell} \bar{\psi}^{m} \psi_{n}.
\end{eqnarray}
The Lagrangian for the $\mathfrak{su} \left( \mathcal{N} \right)$ field $\bm{a}$ includes both a CS term for $d=5$ and a CS term for $d=3$, the latter being suitable multiplied by the field-strength for the $\bm{b}$-field, d$b$.

Explicit expressions for the boundary terms can be also easily obtained replacing Equation~(\ref{q3}) in~(\ref{Bsugra}), in stark contrast with the standard usual case, where multiple, iterative integrations by parts must be performed.

\section{\label{last}Conclusions}

In this paper some general features of transgression forms used as Lagrangians for gauge field theories were briefly reviewed.

A concrete theory is obtained from the general framework by picking a (super)algebra and a symmetric invariant tensor for it. It then becomes important to extract physical information from the Lagrangian. A crucial step in this direction is the separation of the Lagrangian in a way that reflects the inner subspace structure of the gauge algebra. This is especially true in the case of higher-dimensional supergravity, where superalgebras come naturally split into distinct subspaces. Performing this separation using only Leibniz's rule and the definition of a CS form becomes a painstakingly hard task in dimensions higher than three, due to the nonlinearities present in the CS form.

To ease this task we have presented a method, based on the iterative use of the ECHF, which allows one
\begin{enumerate}
\item to separate the Lagrangian in bulk and boundary contributions, and
\item to \emph{easily and systematically} split the Lagrangian in order to
appropriately reflect the subspace structure of the gauge group (as was
illustrated in Section~\ref{su5} by means of an explicit example).
\end{enumerate}

The usefulness of the method has been highlighted by means of an example provided by five-dimensional CS SUGRA. Different scenarios where the Method is applicable are further examined in Ref.~\cite{Iza05} and also, in the context of an eleven-dimensional gauge theory for the M~Algebra, in Ref.~\cite{Iza06c}.

It is interesting to note that, in order to use the Method, one must regard CS forms as a particular case of transgressions. This strongly suggests that they could take central stage on their own~\cite{Bor03,Mor04a,Fat04,Fat05,Bor05,Iza05,Mor05,Mor06a,Mor06b}.

\begin{acknowledgements}
The authors wish to thank J.~A.~de~Azc\'{a}rraga for his warm hospitality at the Universitat de Val\`{e}ncia. F.~I. and E.~R. wish to thank D.~L\"{u}st for his kind hospitality at the Humboldt-Universit\"{a}t zu Berlin and at the Arnold Sommerfeld Center for Theoretical Physics in Munich. P.~S. was partially supported by FONDECYT Grant 1040624 and by Universidad de Concepci\'{o}n through Semilla Grants 205.011.036-1S, 205.011.037-1S. F.~I. and E.~R. were supported by Ministerio de Educaci\'{o}n (Chile) through MECESUP Grant UCO 0209 and by grants from the German Academic Exchange Service (DAAD).
\end{acknowledgements}

\end{document}